\documentclass[aps,prl,twocolumn,superscriptaddress,showpacs]{revtex4}
\usepackage{graphicx}
\usepackage{amsmath}
\usepackage{mciteplus}

\begin{document}
\title{Nonclassical microwave radiation from the dynamical Casimir effect} 
\author{J.R. Johansson}
\email{robert@riken.jp}
\affiliation{Advanced Science Institute, RIKEN, Wako-shi, Saitama, 351-0198 Japan}
\author{G. Johansson}
\affiliation{Microtechnology and Nanoscience, MC2, Chalmers University of Technology, SE-412 96 G{\"o}teborg, Sweden}
\author{C.M. Wilson}
\affiliation{Microtechnology and Nanoscience, MC2, Chalmers University of Technology, SE-412 96 G{\"o}teborg, Sweden}
\author{P. Delsing}
\affiliation{Microtechnology and Nanoscience, MC2, Chalmers University of Technology, SE-412 96 G{\"o}teborg, Sweden}
\author{F. Nori}
\affiliation{Advanced Science Institute, RIKEN, Wako-shi, Saitama, 351-0198 Japan}
\affiliation{Physics Department, The University of Michigan, Ann Arbor, Michigan 48109-1040, USA}
\date{\today}
\begin{abstract}
We investigate quantum correlations in microwave radiation produced by the dynamical Casimir effect in a superconducting waveguide terminated and modulated by a superconducting quantum interference device. We apply nonclassicality tests and evaluate the entanglement for the predicted field states. For realistic circuit parameters, including thermal background noise, the results indicate that the produced radiation can be strictly nonclassical and can have a measurable amount of intermode entanglement. If measured experimentally, these nonclassicalilty indicators could give further evidence of the quantum nature of the dynamical Casimir radiation in these circuits.
\end{abstract}
\maketitle

%
Vacuum fluctuations are fundamental in quantum mechanics, yet they have not so far played an active role in the rapidly advancing field of engineered quantum devices, e.g., for quantum information processing and communication. The main reason being that it has been notably difficult to observe dynamical consequences of the vacuum fluctuations \cite{nation:2012}, let alone use them for applications. The dynamical Casimir effect (DCE) \cite{dodonov:2010, dalvit:2011} is a vacuum amplification process that can produce pairs of photons from quantum vacuum fluctuations by means of nonadiabatic changes in the mode structure of the quantum field, e.g., by a changing boundary condition \cite{moore:1970,fulling:1976} or index of refraction \cite{yablonovitch:1989,uhlmann:2004}. As such it could potentially be applied as a source of entangled microwave photons.

For decades the DCE eluded experimental demonstration, largely due to the challenging prerequisite of nonadiabatic changes in the mode structure with respect to the speed of light. However, using a varying boundary condition in a superconducting waveguide \cite{johansson:2009, johansson:2010}, the experimental observation of the DCE was recently reported \cite{wilson:2011}. This experiment also demonstrated that the dynamical Casimir radiation exhibits the expected two-mode squeezing \cite{dodonov:1990,dodonov:1999,dezael:2010,johansson:2010}, which is a consequence of a nonclassical pairwise photon-creation process. 

The microwave radiation produced by the DCE in superconducting circuits therefore has high potential of being distinctly nonclassical. Whether the state of a quantum field is nonclassical, or if it could be produced by a classical process, may be demarcated by evaluating certain carefully-designed inequalities \cite{miranowicz:2010} for the field observables (nonclassicality tests).  In this paper, we apply such nonclassicality tests to show that the microwave radiation produced by the DCE in these superconducting circuits can be distinctly nonclassical, even when taking into account the background thermal noise \cite{plunien:2000,schutzhold:2002} and higher-order scattering processes. Using auxiliary quantum systems as detectors \cite{dodonov:2012a,dodonov:2012b} could be an alternative to directly measure the field quadratures, which could provide further opportunities to detect nonclassical correlations, e.g., on the single photon-pair level \cite{chen:2011}.

%
{\it DCE in superconducting circuits.---}%
Superconducting circuits are strikingly favorable for amplifying vacuum fluctuations because of their inherently low dissipation, which allows the vacuum state to be reached, and the in-situ tunability of an essential circuit element, namely the Josephson junction (JJ). A JJ is characterized by its Josephson energy, and by arranging two such junctions in a superconducting loop -- a superconducting quantum interference device (SQUID) -- an effective tunable JJ can be produced. The Josephson energy of the effective junction can be tuned by the applied magnetic flux through the SQUID-loop. This in-situ tunability can be used to produce waveguide circuits with tunable boundary conditions \cite{wallquist:2006,laloy:2008,sandberg:2008,yamamoto:2008}, as employed in the DCE experiment in Ref.~\cite{wilson:2011}, and tunable index of refraction \cite{castellanos:2008,nation:2009,latheenmaki:2011}. Tunable JJs are also essential in related DCE proposals based on circuit QED with tunable coupling \cite{deliberato:2009}.

The electromagnetic field confined by a superconducting waveguide, such as a coplanar or strip-line waveguide, can be described quantum mechanically in terms of the flux operator $\Phi(x,t)$. It is related to the voltage operator by $\Phi(x,t) = \int^tdt'V(x,t')$, and to the gauge-invariant superconducting phase operator $\varphi = 2\pi\Phi/\Phi_0$, where $\Phi_0 = h/2e$ is the magnetic flux quantum. The flux field in the transmission line obeys the massless, one-dimensional Klein-Gordon wave equation, $\partial_{xx}\Phi(x,t)-v^{-2}\partial_{tt}\Phi(x,t)=0$, which has independent left- and right-propagating components. Using this decomposition, the field can be written in the form
\begin{eqnarray}
\label{eq:field}
\Phi(x,t) &=& \sqrt{\frac{\hbar Z_0}{4\pi}}\int_{-\infty}^{\infty} \frac{d\omega}{\sqrt{|\omega|}}\times\nonumber\\
&&
\left[a(\omega) e^{-i(-k_\omega x +\omega t)} + b(\omega)e^{-i(k_\omega x +\omega t)}\right],
\end{eqnarray}
where $a(\omega)$ and $b(\omega)$ are the annihilation operators for photons with frequency $\omega/2\pi>0$ propagating to the right (incoming) and left (outgoing), respectively. Here we have used the notation $a(-\omega)=a^\dag(\omega)$, and $k_\omega = \omega/v$ is the wavenumber, $v$ is the speed of light in the waveguide, and $Z_0$ the characteristic impedance.

Using the previously discussed flux-tunable SQUID termination of the waveguide, one can produce a tunable boundary condition (see also Refs.~\cite{alves:2006,silva:2011}) for the quantum field [Eq.~(\ref{eq:field})],
\begin{eqnarray}
\Phi(0,t) + \left.L_{\rm eff}(t)\partial_x\Phi(x,t)\right|_{x=0} = 0,
\end{eqnarray}
that can be characterized by an effective length $L_{\rm eff}(t) = \left(\Phi_0/2\pi\right)^2/(E_J(t)L_0)$, where $L_0$ is the characteristic inductance per unit length of the waveguide and $E_J(t)=E_J[\Phi_{\rm ext}(t)]$ is the flux-dependent effective Josephson energy. To arrive at this boundary condition we have neglected the capacitance of the SQUID and assumed small phase fluctuations, which is justified for a large SQUID plasma frequency \cite{johansson:2009,johansson:2010}. For sinusoidal modulation with frequency $\omega_d/2\pi$ and normalized amplitude $\epsilon$, $E_J(t) = E_J^0 [1 + \epsilon \sin \omega_d t]$, we obtain an effective length modulation amplitude $\delta\!L_{\rm eff} = \epsilon L^0_{\rm eff}$, where $L^0_{\rm eff} = L_{\rm eff}(0)$. A strong modulation (corresponding to an effective velocity $v_{\rm eff}=\delta\!L_{\rm eff}\omega_d$ that is a significant fraction of the speed of light in the waveguide $v$), results in nonadiabatic changes in the mode structure of the quantum field, and the emission of photons as described by the DCE.

The DCE can be analyzed using scattering theory that describes how the time-dependent boundary condition, or region of the waveguide with a time-dependent index of refraction, mixes the otherwise independent left and right propagating modes \cite{lambrecht:1996}. The superconducting circuits considered here were analyzed using this method in Refs.~\cite{johansson:2009,johansson:2010}, where the weak-modulation regime was studied analytically using perturbation theory, and the strong-modulation regime was studied using a higher-order numerical method. 

In the perturbative regime, the resulting output field is correlated at modes with angular frequencies $\omega$ and $\omega_d-\omega$, i.e., symmetrically around half the driving frequency. This intermode symmetry is emphasized when the output field is written for two such correlated modes:
\begin{eqnarray}
\label{eq:output-field-perturbation-simplified-notation}
b_\pm = -a_\pm -i\frac{\delta\!L_{\rm eff}}{v}\sqrt{\omega_+\omega_-}a^\dag_\mp,
\end{eqnarray}
where we have introduced the short-hand notation $a_\pm=a(\omega_\pm)$ and $b_\pm=b(\omega_\pm)$, and where $\omega_\pm = \omega_d/2 \pm \delta\omega$ and $\delta\omega$ is the symmetric detuning. In this perturbation calculation, the small parameter is $\delta\!L_{\rm eff}\sqrt{\omega_-\omega_+}/v \approx \epsilon L_{\rm eff}(0)\omega_d/2v$. Here, even if the input field is in the vacuum state, $\left<a^\dag_\pm a_\pm\right> = 0$, the output field Eq.~(\ref{eq:output-field-perturbation-simplified-notation}) has a nonzero, symmetric photon flux $\left<b^\dag_\pm b_\pm\right> = (\delta\!L_{\rm eff}/v)^2\omega_+\omega_-$, i.e., the dynamical Casimir radiation. Furthermore, the photons in the two modes have bunching-like statistics, where the probability of simultaneously observing one photon in each mode is equal to the probability of observing a photon in one of the modes
$\left<b_+^\dag b_+ b_-^\dag b_-\right> \approx \left<b_\pm^\dag b_\pm\right>$, i.e., they appear in pairs.

For finite temperatures, where thermal noise is present in the input field, and for not so weak modulation, when for example $\delta\!L_{\rm eff}\sqrt{\omega_+\omega_-}/v$ no longer is a small parameter, it is not obvious if or to what extent the above results apply. In these cases there are both classical and nonclassical contributions to the photon flux in the output field, and it becomes necessary to systematically compare the relative importance of such contributions in order to tell if the resulting output field remains nonclassical or not. In the following, we carry out such an analysis using nonclassicality tests and by evaluating the degree of entanglement in the predicted output field.

%
{\it Nonclassicality tests.---}%
The theory of nonclassicality tests has been well developed in quantum optics, and here we briefly review the important results in the notation introduced above for superconducting waveguides. We consider an operator $\hat{f}$ which is defined as a function of the creation and annihilation operators. For the Hermitian operator $\hat{f}^\dagger\hat{f}$ it can then be shown \cite{miranowicz:2010}, using the Glauber-Sudarshan $P$ function formalism, that any classical state of the field satisfies
\begin{equation}
\label{eq:fdf_ineq}
\left<:\hat{f}^\dagger\hat{f}:\right> \geq 0,
\end{equation}
where the condition for classicality that has been used is that the $P$ function must always be non-negative. The $::$ denotes normal ordering.

For the two-mode quadrature-squeezed states that the DCE is known to produce, the natural definition of $\hat{f}$ is
\begin{equation}
\label{eq:f_def}
\hat{f}_\theta = e^{i\theta}\hat{b}_- + e^{-i\theta}\hat{b}_-^\dag +i(e^{i\theta}\hat{b}_+ - e^{-i\theta}\hat{b}_+^\dag),
\end{equation}
where $\theta$ is the angle that defines the principal squeezing axis. With this definition of $\hat{f}_\theta$, a pure two-mode squeezed state is known to violate the inequality (\ref{eq:fdf_ineq}), see, e.g., Ref.~\cite{miranowicz:2010} and references therein. This choice of $\hat{f}_\theta$ is also suitable from an experimental point of view, since $\left<:\hat{f}^\dagger_\theta\hat{f}_\theta:\right>$ can be evaluated from experimentally-accessible quadrature correlations.

We now evaluate the quantum-classical indicator $\left<:\hat{f}^\dagger\hat{f}:\right> = \min\limits_{\theta} \left<:\hat{f}^\dagger_\theta\hat{f}_\theta:\right>$ for the field state produced by the DCE, and discuss the conditions under which this nonclassicality test is violated. For weak driving, using output field Eq.~(\ref{eq:output-field-perturbation-simplified-notation}), and a thermal input field we obtain 
\begin{eqnarray}
 \left<:f^\dag_\theta f_\theta:\right> 
 &=& 
 2(n^{\rm th}_+ + n^{\rm th}_-)\nonumber\\
 &-&
 4 \cos2\theta \frac{\delta L_{\rm eff}}{v}\sqrt{\omega_+\omega_-} (1 + n^{\rm th}_+ + n^{\rm th}_-),
\end{eqnarray}
where $n^{\rm th}_\pm = \left<a_\pm^\dag a_\pm\right> = (\exp(\hbar\omega_\pm/k_BT)-1)^{-1}$ is the thermal photon flux of the input mode with frequency $\omega_\pm$. In this case, $\left<:f^\dag_\theta f_\theta:\right>$ is minimized by taking $\theta = 0$, and it is negative if $(\delta L_{\rm eff}/v)\sqrt{\omega_+\omega_-} \gtrsim (n^{\rm th}_+ + n^{\rm th}_-)/2$, or, equivalently, $\epsilon\gtrsim 2v/(L^0_{\rm eff}\omega_d)(n^{\rm th}_++n^{\rm th}_-)/2$. This indicates that the field state in the form Eq.~(\ref{eq:output-field-perturbation-simplified-notation}) is distinctly nonclassical for a vacuum input field, and potentially also for low-temperature thermal input fields. 

\begin{figure}[t]
\begin{center}
\includegraphics[width=8.5cm]{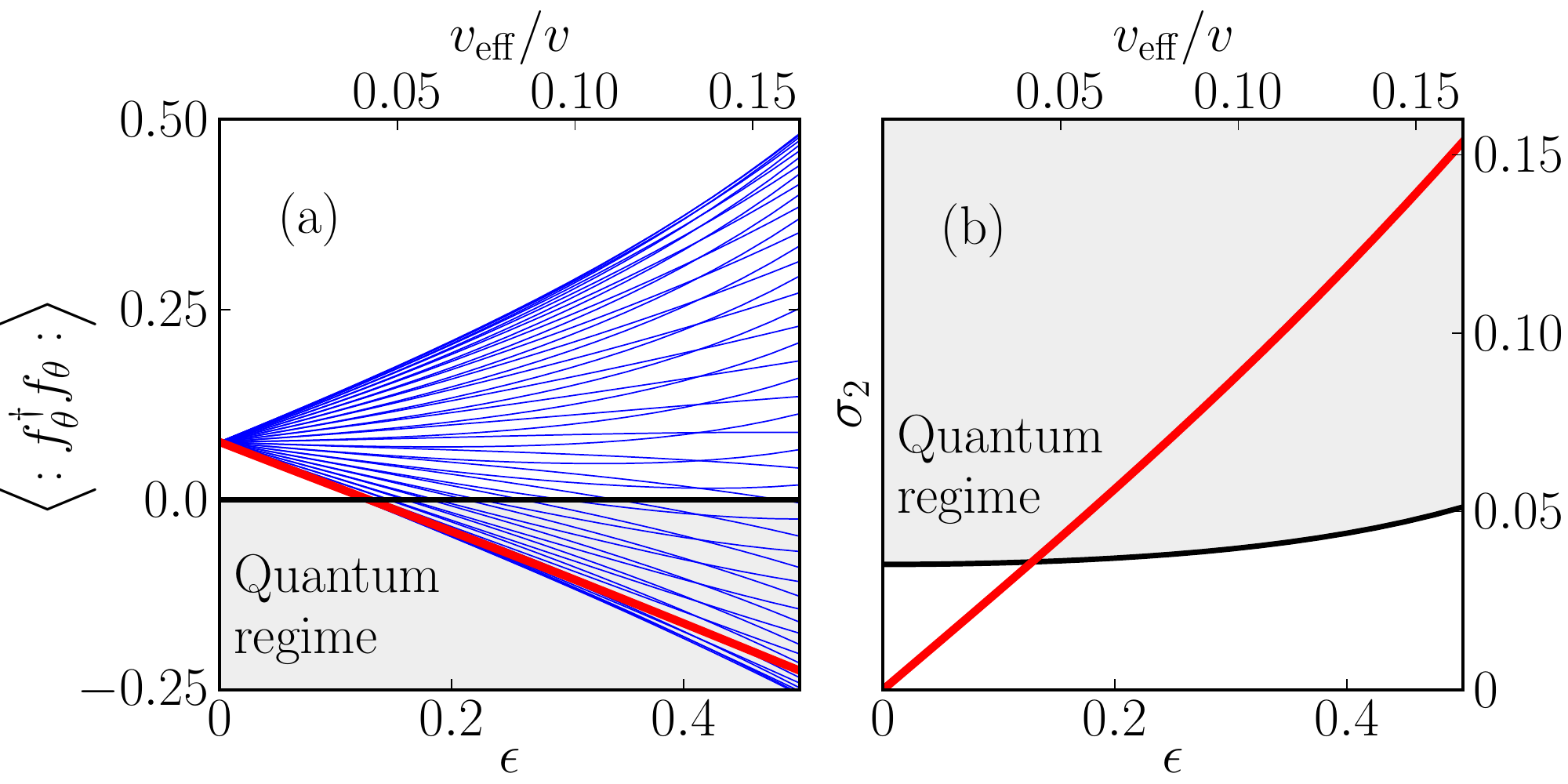}
\caption{
(color online) (a) The quantum-classical indicator $\left<:f^\dag_\theta f_\theta:\right>$ as a function of driving amplitude $\epsilon$ for a range of $\theta$ values in the interval $[0,2\pi]$ (blue), and for $\theta = 0$ (red), which is the optimal $\theta$ in the perturbation regime. Due to the thermal input field, $\left<:f^\dag_\theta f_\theta:\right> > 0$ for small $\epsilon$. However, when $\epsilon$ is sufficiently large $\left<:f^\dag f:\right> < 0$, which conclusively rules out that the field state is of classical origin. (b) The two-mode squeezing $\sigma_2$ as a function of the dimensionless driving amplitude $\epsilon$ (red), together with the right-hand side of Eq.~(\ref{eq:sigma-2-ineq}) (black), which defines the boundary between the classical and quantum regions.
Parameters: $\omega_d/2\pi = 10$ GHz, $\delta\omega/\omega_d = 0.15$, $T = 50$ mK. Other parameters are from Ref.~\cite{wilson:2011}.}
\label{fig:dce-fdf-range-and-sigma2-vs-eps}
\end{center}
\end{figure}

To investigate whether the nonclassical characteristics of the DCE radiation remain for realistic input field temperatures and when the driving amplitude is increased beyond the perturbative regime, we also evaluate $\left<:f^\dag_\theta f_\theta:\right>$ by solving the scattering problem numerically. The results of this calculation are presented in Fig.~\ref{fig:dce-fdf-range-and-sigma2-vs-eps}(a), showing that for sufficiently large driving amplitude $\left<:f^\dag f:\right> < 0$ even at typical temperatures for superconducting circuits, and including higher-order scattering processes. We therefore conclude that nonclassical characteristics of the DCE radiation can be sufficiently robust to remain important in realistic experimental situations. Evaluating $\left<:f^\dag f:\right>$ from experimentally measured field quadratures therefore appears be a viable method to conclusively demonstrate the quantum statistics of the dynamical Casimir radiation.

%
{\it The nonclassicality test in terms of $\sigma_2$.---}%
To further relate to the experimental demonstrations of the DCE, it is instructive to formulate the nonclassicality test in terms of the two-mode squeezing $\sigma_2$, which was measured in Ref.~\cite{wilson:2011}. The two-mode squeezing is defined as $\sigma_2 = (\left<I_-I_+\right> - \left<Q_-Q_+\right>)/\left((\left<I_-^2\right>+\left<I_+^2\right>+\left<Q_-^2\right>+\left<Q_+^2\right>)/2\right)$, where $I_\pm = \left(\hbar\omega_\pm Z_0/8\pi\right)^{1/2}\left(e^{i\phi}b_\pm +e^{-i\phi}b_\pm^\dag\right)$ and $Q_\pm = -i\left(\hbar\omega_\pm Z_0/8\pi\right)^{1/2}\left(e^{i\phi} b_\pm - e^{-i\phi} b_\pm^\dag\right)$ are the voltage quadratures. Using this expression for $\sigma_2$, we can write the inequality $\left<:f^\dag_\theta f_\theta:\right> < 0$ as
\begin{eqnarray}
\label{eq:sigma-2-ineq}
\sigma_2
  &>& 
\frac{2\sqrt{\omega_+\omega_-}\left(n_+ + n_-\right)}
{\omega_+ \left[2n_+ + 1\right] + \omega_-\left[2n_- + 1\right]},
\end{eqnarray}
where $n_\pm = \left<b_\pm^\dag b_\pm\right>$ is the photon flux (thermal and DCE) for the output mode with frequency $\omega_\pm$, and where we have taken $\theta=\phi+\pi/4$ to relate $\sigma_2$ and $\left<:f^\dag_\theta f_\theta:\right>$.

Equation (\ref{eq:sigma-2-ineq}) suggests that a non-zero two-mode squeezing does not necessarily imply that the field is a strictly nonclassical state [by the criterium of Eq.~(\ref{eq:fdf_ineq}) and the current definition of the operator $\hat{f}$]. However, if the magnitude of the two-mode squeezing exceeds the right-hand side of Eq.~(\ref{eq:sigma-2-ineq}), the field is {\it guaranteed} to be distinctively nonclassical (i.e., squeezed vacuum rather than a squeezed thermal state). Since the expectation values in the right-hand side of Eq.~(\ref{eq:sigma-2-ineq}) can be measured experimentally, this could be a practical formulation for the experimental evaluation of the nonclassicality test. 

Figure \ref{fig:dce-fdf-range-and-sigma2-vs-eps}(b) shows the two-mode squeezing together with the boundary between the classical and quantum regimes, as defined by Eq.~(\ref{eq:sigma-2-ineq}). With the parameters used in Fig.~\ref{fig:dce-fdf-range-and-sigma2-vs-eps}, the boundary corresponds to the squeezing $\sigma_2 \approx 0.04$. Experimental measurements \cite{wilson:2011} have demonstrated significantly larger squeezing for the dynamical Casimir radiation, but at the same time the measured photon flux was larger than in the current calculations due to the presence of low-$Q$ resonances in the transmission line. An increased photon flux increases the value of the boundary in Eq.~(\ref{eq:sigma-2-ineq}) and makes the violation of the inequality more demanding. However, by reducing the driving strength to get a lower photon flux a violation of the nonclassicality test Eq.~(\ref{eq:sigma-2-ineq}) should be achievable with an experimental setup like the one in Ref.~\cite{wilson:2011}, although increased measurement time and averaging may be necessary to obtain sufficient sensitivity.

%
{\it Entanglement.---}%
The two-mode squeezing and the nonclassicality tests discussed above demonstrate that the DCE radiation is nonclassical. The quantum nature of the radiation originates from the entanglement in {\it individual pairs of photons}. To quantify the entanglement between two {\it entire modes} with frequencies adding up to the driving frequency, we evaluate the logarithmic negativity $\mathcal{N}$ \cite{adesso:2007}, which is an entanglement measure for Gaussian states that is frequently used in quantum optics, and recently also in microwave circuits \cite{flurin:2012} and nanomechanical systems \cite{joshi:2012}. The logarithmic negativity is positive for entangled states, and it can be calculated from the covariance matrix $V_{\alpha\beta} = \frac{1}{2}\left<R_\alpha R_\beta+R_\beta R_\alpha\right>$, where $R^{\rm T} = \left(q_-, p_-, q_+, p_+\right)$ is a vector with the quadratures as elements: $q_\pm = (b_\pm + b_\pm^\dag)/\sqrt{2}$ and $q_\pm = -i(b_\pm - b_\pm^\dag)/\sqrt{2}$. 

\begin{figure}[t]
\begin{center}
\includegraphics[width=8.5cm]{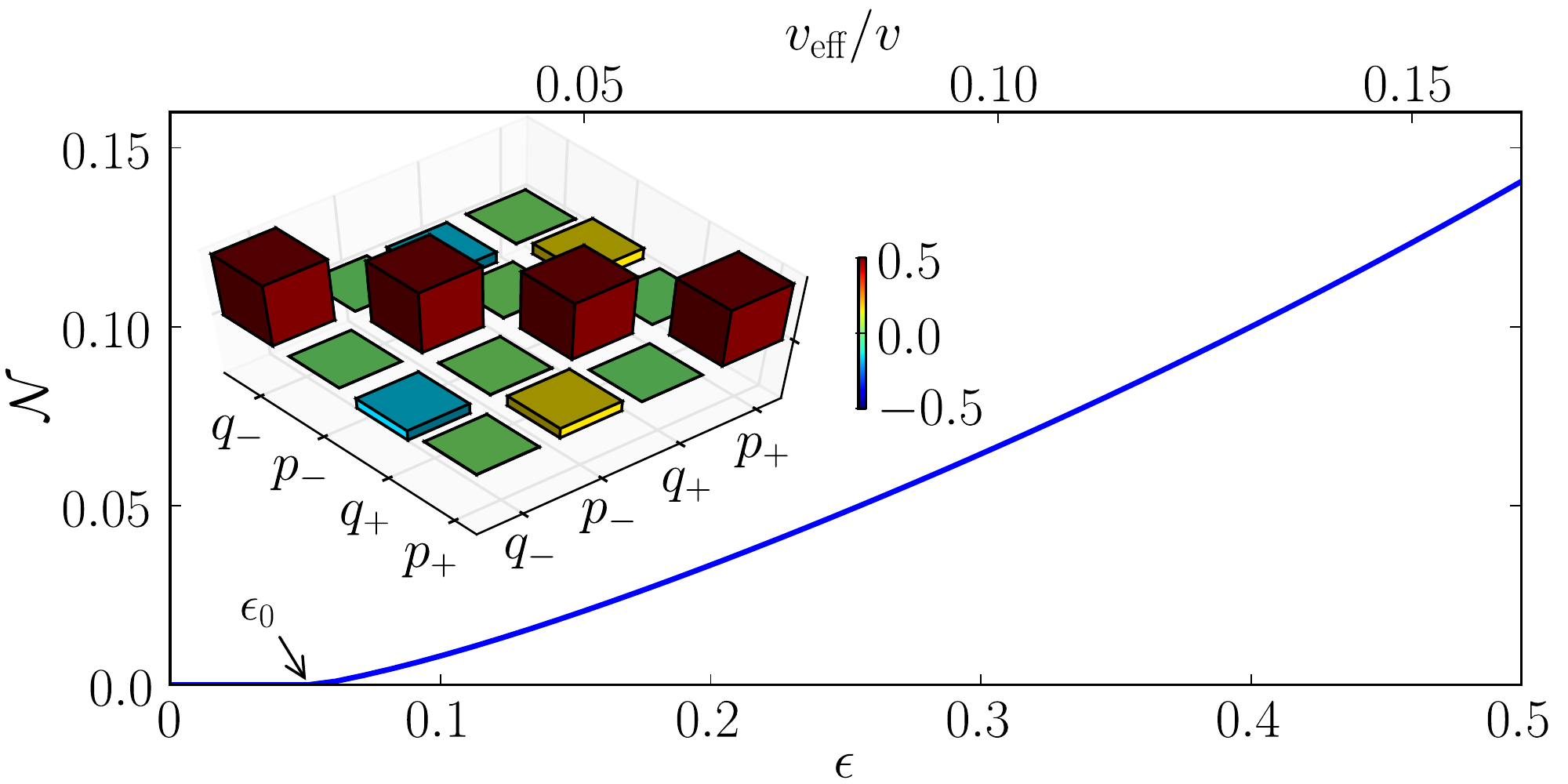}
\caption{(color online) The logarithmic negativity $\mathcal{N}$ as a function of the normalized modulation amplitude $\epsilon$. The onset of nonzero $\mathcal{N}$ is $\epsilon_0$. The parameters are the same as in Fig.~\ref{fig:dce-fdf-range-and-sigma2-vs-eps}. Inset: The covariance matrix for $\epsilon = 0.5$. The diagonal quadrature correlations correspond to the vacuum fluctuations and the photon flux due to the DCE and of thermal origin. The nonzero, off-diagonal elements correspond to the two-mode correlations produced by the DCE.}
\label{fig:dce-log-neg-vs-eps}
\end{center}
\end{figure}

The covariance matrix can be evaluated both analytically and numerically, and also constructed from experimental quadrature measurements. The numerically calculated covariance matrix is shown in the inset in Fig.~\ref{fig:dce-log-neg-vs-eps} for typical parameters. Given the covariance matrix for the two selected modes, it is straightforward to evaluate the logarithmic negativity, defined as $\mathcal{N} = \max[0, -\log(2\nu_-)],$ where $\nu_- = (\sigma/2-(\sigma^2 - 4 \det V)^{1/2}/2)^{1/2},$ and $\sigma = \det A + \det B -2 \det C,$ where the $A, B$, and $C$ are $2\times2$ submatrices of the covariance matrix $V = \left(A, C; C^T, B\right)$.

\begin{figure}[t]
\begin{center}
\includegraphics[width=8.5cm]{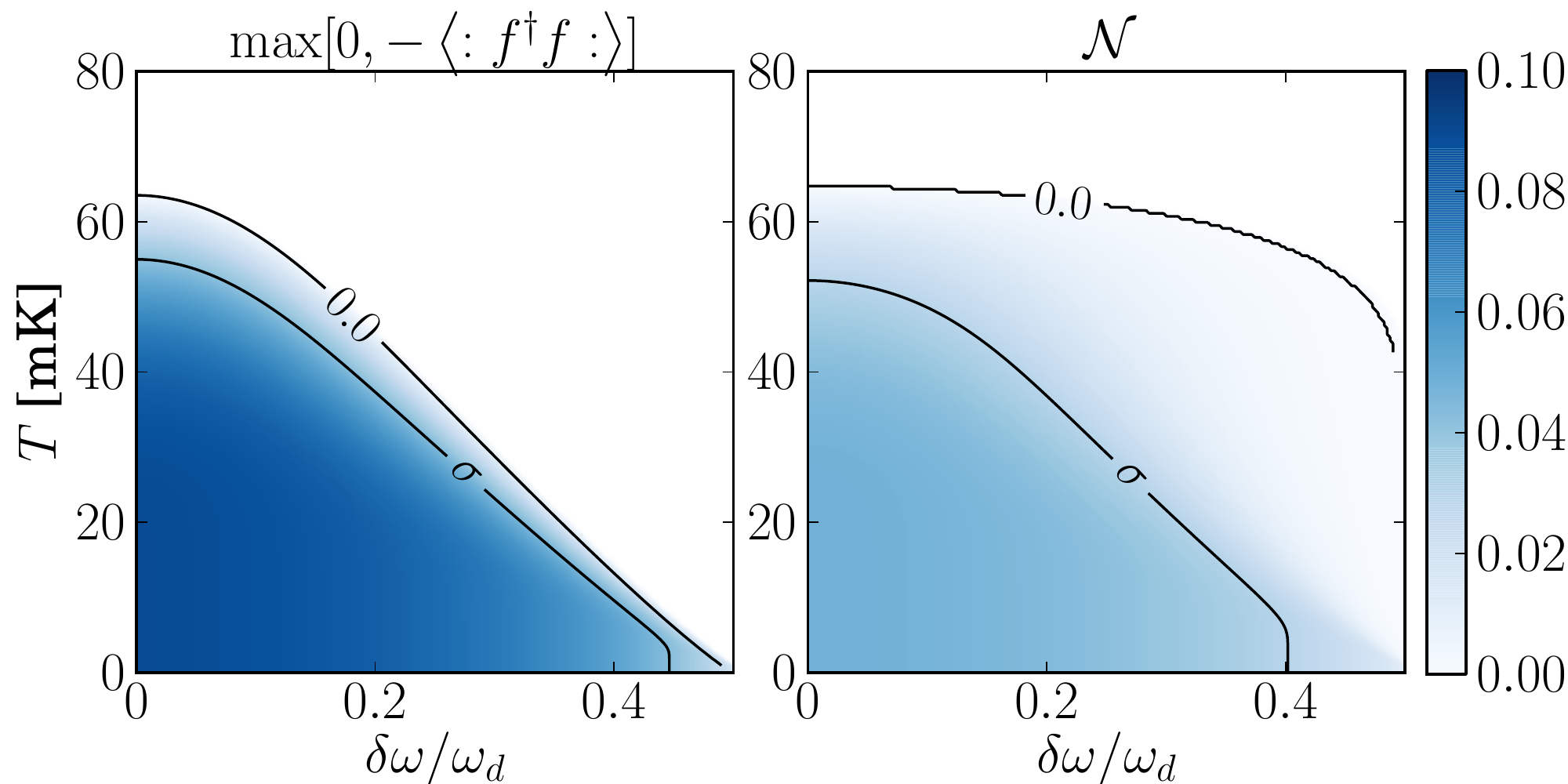}
\caption{(color online) The region of nonclassical radiation (blue), visualized using $-\left<:f^\dag f:\right>$ (left) and the logarithmic negativity $\mathcal{N}$ (right), as a function of the temperature $T$ and the detuning $\delta\omega$, for $\epsilon=0.15$ and other parameters as in Fig.~\ref{fig:dce-fdf-range-and-sigma2-vs-eps}. Although the nonclassical region is larger for $\mathcal{N}$ than for $\left<:f^\dag f:\right>$, $\mathcal{N}$ is small in the region where $\left<:f^\dag f:\right>$ is non-negative (white), and the regions where the measures violate classicality with a one-$\sigma$ confidence are quite similar.}
\label{fig:combined-qci-region}
\end{center}
\end{figure}

The logarithmic negativity for the DCE (see also Ref.~\cite{guerreiro:2012}) is shown in Fig.~\ref{fig:dce-log-neg-vs-eps}. At zero temperature and small drive amplitudes, it is proportional to the driving amplitude $\mathcal{N} = \epsilon L^0_{\rm eff}\omega_d/v$. For finite temperatures and small detuning $\delta\omega$, the onset of nonzero logarithmic negativity is at $\epsilon_0 \approx 2v/(L^0_{\rm eff}\omega_d)(n^{\rm th}_+n^{\rm th}_-)^{1/2}$, after which it increases with the driving amplitude. For sufficiently large driving amplitude, $\epsilon \gtrsim 0.06$, the quantum correlations overcome the thermal noise and the two matching output modes are entangled. Comparing Figs.~\ref{fig:dce-fdf-range-and-sigma2-vs-eps} and \ref{fig:dce-log-neg-vs-eps} implies that the logarithmic negativity is a stronger indicator of the nonclassicality of the field state than the inequality (\ref{eq:fdf_ineq}) with our definition of $\hat{f}$. This is also shown in Fig.~\ref{fig:combined-qci-region}, which visualizes the nonclassical regions as a function of temperature and detuning, as well as the sensitivity to uncorrelated classical quadrature noise introduced in the detector (the one-$\sigma$ contour line). However, when taking this sensitivity into consideration, the two measures appear to be of similar practical usefulness.

%
{\it Conclusion.---}%
We have theoretically investigated quantum correlations in the radiation produced by the DCE in a superconducting waveguide by evaluating nonclassicality tests and the logarithmic negativity. These measures indicates that the devices used in Ref.~\cite{wilson:2011}, should have access to regimes where the produced radiation is strictly nonclassical. We have formulated practical inequalities with experimentally obtainable observables that could be used to directly verify the quantum nature of the measured radiation in future DCE experiments. We also note that recently two-mode squeezed states have been generated in microwave circuits using other mechanisms, for example parametric amplification using the nonlinear response \cite{castellanos:2008,eichler:2011} or time-varying index of refraction \cite{latheenmaki:2011} of SQUID arrays and JJs \cite{bergeal:2010,flurin:2012}. The nonclassicality tests discussed here could also be applied to analyze the radiation produced in these experiments. We believe that a demonstration of a nonclassicality violation in superconducting circuits, or other promising systems \cite{braggio:2005,naylor:2009,faccio:2011,carusotto:2012}, could pave the way to the experimental exploration of the continuous production of entangled microwave photons by the DCE, and possible applications thereof in, for example, quantum information processing \cite{you:2011,you:2005,buluta:2011}. As such it could become a novel practical application of microwave quantum vacuum fluctuations.

%

JRJ is supported by the JSPS Foreign Postdoctoral Fellowship No. P11501.
GJ, CMW and PD acknowledge financial support from the Swedish Research Council, the European Research Council as well as the European Comission through the FET Open project PROMISCE.
FN is partially supported by the ARO, NSF No. 0726909, JSPS-RFBR No. 12-02-92100, Grant-in-Aid for Scientific Research (S), MEXT Kakenhi on Quantum Cybernetics, and the JSPS-FIRST program. 

%
\bibliography{references}
\end{document}